\documentclass[]{llncs}
\usepackage[T1]{fontenc}
\usepackage{amssymb}
\usepackage{amsmath}
\usepackage{algorithmicx}
\usepackage{algorithm}
\usepackage{algpseudocode}
\usepackage[misc]{ifsym}

\begin{document}
\title{An FPT Constant-Factor Approximation Algorithm for Correlation Clustering\thanks{The work has been supported by the National Natural Science Foundation of China (No. 61772314, 61761136017, and 62072275).}}
\titlerunning{An FPT Constant-Factor Approximation for Correlation Clustering}

\author{Jianqi Zhou \and
	Zhongyi Zhang \and
	Jiong Guo$^{(\textrm{\Letter})}$
	} 
\authorrunning{J. Zhou et al.}
% First names are abbreviated in the running head.
% If there are more than two authors, 'et al.' is used.
%
\institute{School of Computer Science and Technology, Shandong University, Qingdao, China \\
	\email{jqzhou@mail.sdu.edu.cn,zhangzhongyi@mail.sdu.edu.cn,jguo@sdu.edu.cn}}

\maketitle{}% typeset the header of the contribution

\begin{abstract}

	The Correlation Clustering problem is one of the most extensively studied clustering formulations due to its wide applications in machine learning, data mining, computational biology and other areas.
    We consider the Correlation Clustering problem on general graphs, where given an undirected graph (maybe not complete) with each edge being labeled with $\langle + \rangle$ or $\langle - \rangle$, the goal is to partition the vertices into clusters to minimize the number of the disagreements with the edge labeling: the number of $\langle - \rangle$ edges within clusters plus the number of $\langle + \rangle$ edges between clusters.
    Hereby, a $\langle + \rangle$ (or $\langle - \rangle$) edge means that its end-vertices are similar (or dissimilar) and should belong to the same cluster (or different clusters), and  ``missing'' edges are used to denote that we do not know if those end-vertices are similar or dissimilar.

    Correlation Clustering is NP-hard, even if the input graph is complete, and Unique-Games hard to obtain polynomial-time constant approximation on general graphs.
    With a complete graph as input, Correlation Clustering admits a $(1.994+\varepsilon )$-approximation.
    We investigate Correlation Clustering on general graphs from the perspective of parameterized approximability.
    We set the parameter $k$ as the minimum number of vertices whose removal results in a complete graph, and obtain the first FPT constant-factor approximation for Correlation Clustering on general graphs which runs in $2^{O(k^3)} \cdot \textrm{poly}(n)$ time.
	
	%The abstract should briefly summarize the contents of the paper in 150--250 words.
	
	\keywords{  Cluster editing  \and Clustering with partial information \and Parameterized approximation \and   Incomplete graphs.}
\end{abstract}

\section{Introduction}
\label{sec:Introduction}

Due to wide applications in machine learning, data mining, computational biology, and other areas, the Correlation Clustering problem, introduced by Bansal, Blum, and Chawla~\cite{DBLP:journals/ml/BansalBC04}, has been extensively studied concerning its computational complexity, approximability, and parameterized complexity. 
Bansal, Blum, and Chawla~\cite{DBLP:journals/ml/BansalBC04} initially defined the Correlation Clustering problem on complete graphs as follows: given a complete, undirected graph of $n$ vertices (items), where each edge is labeled either $\langle + \rangle$ or $\langle - \rangle$ depending on whether its two end-vertices are considered to be similar or dissimilar, the goal is to produce a partition of the vertices (a clustering) that agrees as much as possible with the edge labels.
That is, we seek for a clustering that maximizes the sum of the number of $\langle + \rangle$ edges within clusters and the number of $\langle - \rangle$ edges between clusters (equivalently, minimizes the number of disagreements: the number of $\langle - \rangle$ edges inside clusters plus the number of $\langle + \rangle$ edges between clusters).
Note that the Correlation Clustering problem does not specify the number of clusters in advance.

Correlation Clustering with minimizing disagreements and with maximizing agreements are equivalent with respect to classical complexity, and NP-complete on complete graphs~\cite{DBLP:journals/ml/BansalBC04}.
However, they differ in approximability.
For the version of minimizing disagreements on complete graphs, Bansal, Blum and Chawla~\cite{DBLP:journals/ml/BansalBC04} gave the first constant-factor approximation algorithm.
Charikar, Guruswami and Wirth~\cite{CHARIKAR2005360} improved the approximation factor to~4 by rounding the standard linear programming (LP) relaxation and proved APX-hardness, meaning that the problem does not admit a polynomial-time approximation scheme (PTAS). 
Ailon, Charikar and Newman~\cite{10.1145/1411509.1411513} proposed a combinatorial randomized 3-approximation algorithm and a LP-based randomized 2.5-approximation  algorithm.
Then, Zuylen and Williamson~\cite{DBLP:journals/mor/ZuylenW09} gave a deterministic 3-approximation algorithm by following the paradigm of Ailon et al.~\cite{10.1145/1411509.1411513} and using an LP relaxation.
Finally, the LP-based pivoting algorithm was further improved to a 2.06-approximation by Chawla et al.~\cite{10.1145/2746539.2746604} and to a $(1.994+ \varepsilon )$-approximation by Cohen{-}Addad, Lee and Newman~\cite{9996889}.
For the version of maximizing agreements on complete graphs,
Correlation Clustering admits a PTAS~\cite{DBLP:journals/ml/BansalBC04}.

However, in most real-world applications, we are not able to obtain the similarity information of all item pairs, which results in incomplete graphs as input for Correlation Clustering. 
A missing edge means that its two end-vertices might be clustered together or be separated in different clusters. 
With missing edges allowed, we arrive at the Correlation Clustering on general graphs. 
For the version of minimizing disagreements on general graphs,
Charikar, Guruswami and Wirth~\cite{CHARIKAR2005360} and Demaine et al.~\cite{DEMAINE2006172} independently proposed $O(\log n)$-approximation even with positively weighted edges.
They also proved Correlation Clustering on general weighted (or unweighted) graphs is equivalent to Minimum Multicut on weighted (or unweighted) graphs~\cite{CHARIKAR2005360,DEMAINE2006172}, and thus obtaining polynomial-time constant-factor approximation is Unique-Game hard~\cite{DBLP:journals/cc/ChawlaKKRS06}.
For the version of maximizing agreements on general edge-weighted graphs,
Swamy~\cite{10.5555/982792.982866} and Charikar, Guruswami and Wirth~\cite{CHARIKAR2005360} independently derived a 0.766-approximation by rounding semidefinite programs.

\textbf{Our contributions.}
We focus on the minimizing disagreements version on general unweighted graphs.
As mentioned above, obtaining a polynomial-time constant-factor approximation for this version is Unique-Games hard~\cite{CHARIKAR2005360,DBLP:journals/cc/ChawlaKKRS06,DEMAINE2006172}.
We aim at designing an FPT constant-factor approximation algorithm.
An FPT $c$-approximation algorithm returns a $c $-approximative solution in $f(k)\cdot \vert I \vert^{O(1)}$ time, where $f$ is a computable function and $k$ is a parameter of the input~$I$.
Motivated by the polynomial-time inapproximability and the $(1.994+ \varepsilon )$-approximation algorithm~\cite{9996889} for Correlation Clustering on complete unweighted graphs, we use the so-called “distance from approximability” concept for designing FPT approximation algorithms, that is, introducing a parameter measuring the distance between the inapproximable and approximable cases and designing approximation algorithms with the exponential running time restricted to the parameter.
This concept has been used to design FPT approximation algorithms for TSP on general graphs~\cite{zhou_et_al:LIPIcs.ISAAC.2022.50}. 
Here, we use a structure parameter $k$ to measure the distance of a general graph from the complete graph.
We choose $k$ as the minimum number of vertices whose removal results in a complete graph, in other words, that is the minimum number of vertices to cover the missing edges.
Given a general instance, we can apply the search tree algorithms for Verter Cover~\cite{downey2013fundamentals,niedermeier2006invitation} to determine the value of $k$ in $O(2^{k}\cdot n^2)$ time.
Using $k$ as parameter, we achieve a $(\frac{18}{\delta^2}+7.3)$-approximation for Correlation Clustering on general unweighted graphs with a running time of $2^{O(k^3)} \cdot \text{poly}(n)$, where $\delta =\frac{1}{65}$.
To our best knowledge, this is the first FPT constant-factor approximation algorithm for Correlation Clustering on general graphs.
Due to lack of space, most proofs are omitted.

\section{Preliminaries}

We introduce some basic definitions and notations.
We consider an undirected graph $G = (V,E)$ on $n$ vertices and
let $e(u,v)$ denote the label $\langle + \rangle$ or $\langle - \rangle$ of the edge $(u,v)$.
We call an edge $(u, v)$ ``positive'' if $e(u,v)=\langle +\rangle$ and ``negative'' if $e(u,v)=\langle - \rangle$.
If $(u,v) \notin E$, we call $(u,v)$ an ``empty'' edge.
Let $E^{\langle + \rangle}=\{(u,v) \mid e(u,v)=\langle + \rangle\}$ denote the set of all positive edges, and let $G^{\langle + \rangle}=(V, E^{\langle + \rangle})$ be the graph induced by $E^{\langle + \rangle}$.
Likewise we define $E^0$ and $G^0$ for empty edges.

Let $N^+ (u)=\{v \mid  e(u,v)=\langle + \rangle\}$, $N^- (u)=\{v \mid  e(u,v)=\langle - \rangle\}$ and $N^0 (u)=\{v \mid  (u,v) \notin E\}$ denote the set of the positive, negative and empty neighbors of $u$, respectively.
In addition, $N^+ [u]=N^+ (u) \cup \{u\}$.
For a set of vertices $V^\prime \subseteq V$, set $N^+[V^\prime]=\cup_{v \in V^\prime} N^+[v]$ and let $N^+(V^\prime)= N^+[V^\prime]\setminus V^\prime$ denote the set of the positive neighbors of $V^\prime$.
Analogously, we define $N^-(V^\prime)$ and $N^0(V^\prime)$ for negative and empty neighbors of $V^\prime$.
For two sets of vertices $V_1,V_2$, let $E^+(V_1,V_2)=\{(u,v)\in E^{\langle + \rangle}\mid u\in V_1, v\in V_2 \}$ denote the set of $\langle + \rangle$ edges between $V_1$ and $V_2$, and $E^-(V_1,V_2)=\{(u,v)\in E^{\langle - \rangle}\mid u\in V_1, v\in V_2 \}$ denote the set of $\langle - \rangle$ edges between $V_1$ and $V_2$.
For a graph $G=(V,E)$ and a set of vertices $V^\prime \subseteq V$, let $\overline{V^\prime} $ denote the complement $V\setminus V^\prime$ and $G-V'$ denote the graph induced by $V\setminus V'$.
For a positive integer $i$, set $[i]=\{ 1,\cdots,i\}$.

In this paper, we focus on the Correlation Clustering problem of minimizing disagreements on general unweighted graphs. 
We use $\mathcal{C}$ to denote a clustering, $C$ to denote a cluster in a clustering, and OPT($G$) to denote an optimal clustering of $G$.
In general, for a clustering $\mathcal{C} $, let $\mathcal{C}(v)$ denote the set of vertices in the same cluster as $v$.
We call an edge $(u,v)$ a ``positive mistake'' if $e(u,v)=\langle + \rangle$ and $u\notin \mathcal{C}(v)$.
We call an edge $(u,v)$ a ``negative mistake'' if $e(u,v)=\langle - \rangle$ and $u\in \mathcal{C}(v)$.
The total mistakes of a clustering $\mathcal{C}$ in $G$ is the sum of positive and negative mistakes made by $\mathcal{C}$, denoted by $m_G(\mathcal{C})$.
When there is no ambiguity, we may omit the subscript.

\section{An FPT Constant-Factor Approximation Algorithm}

We consider a Correlation Clustering instance where the input graph can be transformed into a complete graph by removing at most $k$ vertices, in other words,
at most $k$ vertices covering all empty edges.
Computing these $k$ vertices is equivalent to solving the Vertex Cover problem of the graph $G^0$,  which can be done in $O(2^{k}\cdot n^2)$ time by a simple search tree~\cite{downey2013fundamentals,niedermeier2006invitation}.
The basic idea is that one end-vertex of every empty edge has to be added to this set of vertices.
We use the ``bad'' vertices to refer to the $k$ vertices and use $V^b$ to denote the set of all bad vertices, $\vert V^b \vert \leq k$.
The remaining vertices are called ``good'' and we use $V^g$ to denote the set of all good vertices.

The subgraph induced by each subset of $V^g$ is complete, and we can use the polynomial-time $(1.994+ \varepsilon )$-approximation algorithm~\cite{9996889} to find a good clustering of this complete subgraph.
Thus, the difficulty lies in how to produce the clusters containing bad vertices, and how to bound mistakes caused by these clusters. 

The high-level idea of our algorithm is as follows.
First, we guess the partition of bad vertices in an optimal clustering.
Second, we separate the bad vertices in different clusters by a multiway cut of an auxiliary graph and obtain some ``$\langle + \rangle$-edge-connected'' subgraphs, each of which only contains the bad vertices from the same subset of the partition.
Then, we distinguish several cases to produce a not-bad clustering for each subgraph and merge them together to an overall solution. 

\subsection{Separate Bad Vertices in Different Clusters}

The first step of our algorithm ``guesses'' the partition of bad vertices in an optimal clustering OPT.
This can be done by enumerating all possible partitions of bad vertices, generating a clustering for each partition, and returning the best one of all these clusterings. 
In each partition case, we analyze the performance of the clustering generated by our algorithm by comparing it with the optimal solution which can be achieved under the assumption that the bad vertices have to be clustered according to the partition, that is, the bad vertices in one subset of the partition being in the same cluster and bad vertices in different subsets ending up in different clusters. 
 
Consider one partition case, where the bad vertices are partitioned into $k^\prime$ subsets $B_1,\cdots, B_{k^\prime}$ where $k^\prime \leq k$.
Then the optimal solution OPT for this case contains  $k^\prime$ disjoint clusters $C_1^{\text{opt}}, \cdots,C_{k^\prime}^{\text{opt}} $ with $C_i^{\text{opt}} \supseteq B_i$ for each $i \in [k^\prime]$ and other clusters of OPT contain only good vertices.
Note that the $\langle + \rangle$ edges between different clusters are the positive mistakes in OPT.
In other words, if we remove the $\langle + \rangle$ edges between different clusters, then for $i\neq j$, $C_i^{\text{opt}}$ and $C_j^{\text{opt}}$ are disconnected in $G^{\langle + \rangle}$, and thus, $B_i$ and $B_j$ are disconnected too.
Thus, if we contract $B_i$ into a vertex $b_i$ in $G^{\langle + \rangle}$ for each $i \in [k^\prime]$, the set of all positive mistakes in OPT is a multiway cut for $k^\prime$ terminals $b_1,\cdots, b_{k^\prime}$.
Here, we recall the definition of the Multiway Cut problem:
given a graph $G=(V,E)$ with edge weights $w: E \rightarrow \mathbb{R} ^+$ and $k^\prime$ terminals $t_1, \cdots, t_{k^\prime}$, find a minimum-weight subset of edges $F \subseteq  E$ such that no pair of terminals is connected in $(V, E\backslash F)$.
This observation motivates us to construct an auxiliary graph $H^{\langle + \rangle}$ from $G^{\langle + \rangle}$ by contracting $B_i$ into a vertex $b_i$,
using the weight of edge $(b_i, v)$ to represent the number of $\langle + \rangle$ edges between $B_i$ and $v$ for all $v \notin B_i$, and setting the weight of other $\langle + \rangle$ edges connecting two good vertices to be 1.
More details are illustrated in Algorithm \texttt{CC}.
We conclude that the minimum weight of the multiway cut of the auxiliary graph $H^{\langle + \rangle}$ with $b_1,\cdots, b_{k^\prime}$ as terminals is a lower bound for $m(\text{OPT})$, the number of mistakes of OPT.

Then we use the polynomial-time 1.2965-approximation algorithm~\cite{10.1145/2591796.2591866} to get a multiway cut on $H^{\langle + \rangle}$, and remove all $\langle + \rangle$ edges in the multiway cut from the original graph $G$.
The number of removed the $\langle + \rangle$ edges is at most $1.3\cdot m(\text{OPT})$.
Afterwards, we obtain $k^\prime$ $\langle + \rangle$-edge-connected subgraphs $G_1, \cdots, G_{k^\prime}$ containing bad vertices $B_1, \cdots, B_{k^\prime}$, respectively, and one subgraph $G_{k^\prime +1}$ containing only good vertices. 
Since there are no $\langle + \rangle$ edges left between different subgraphs, we only need to find a good clustering $\mathcal{C}_i$ for each subgraph $G_i$ for $i\in [k^\prime]$, and then merge $\mathcal{C}_1,\cdots, \mathcal{C}_{k^\prime}$ and the clustering $\mathcal{C}_{k^\prime +1}$ computed for the complete subgraph $G_{k^\prime +1}$ to a clustering as output.
We use Algorithm \texttt{CC} to denote the overall framework of our algorithm and introduce the details of each step later. 
\newline

\noindent\textbf{Algorithm \texttt{CC}}: an algorithm for Correlation Clustering on general graphs

Input: a graph $G$ with edge labeling.

Output: a clustering of $G$.
\begin{enumerate}
    \item Compute the minimum vertex cover $V^b$ of the graph $G^0=(V, E^0)$ by the search tree technique. Set $ k=\vert V^b\vert$.
    \item Enumerate all possible partitions of bad vertices in $V^b$ into $k^\prime$ subsets for each $k^\prime \in [k]$. 
    For each $k^\prime$-partition, do the following:
    \begin{enumerate}
		\item Construct an auxiliary graph $H^{\langle + \rangle}$ from $G^{\langle + \rangle}$ as follows: set the weight of the $\langle + \rangle$ edges connecting two good vertices to be 1; for each $i \in[k^\prime]$, contract $B_i$ into a vertex $b_i$, and set the weight of the edge $(b_i, v)$ to be the number of $\langle + \rangle$ edges between $B_i$ and $v$ for each $v \in V^g$; for each $i,j\in [k^\prime]$ with $i\neq j$, set the weight of $(b_i, b_j)$ to be the number of  $\langle + \rangle$ edges between $B_i$ and $B_j$.
		\item Compute a multiway cut $F$ for $k^\prime$ terminals $b_1,\cdots, b_{k^\prime}$ on the graph $H^+$ by using Sharma and Vondr\'{a}k's algorithm~\cite{10.1145/2591796.2591866}.
		\item Construct an auxiliary graph $H$ from $G$ as follows: remove all $\langle + \rangle$ edges between $B_i$ and $B_j$ for $i \neq j$;
		remove all $(u,v)\in F$ with $u\in V^g$ and $v \in V^g$;
		for each $i\in [k^\prime]$ and each $v\in V^g$, if $(v,b_i) \in F$, remove all $\langle + \rangle$ edges between $v$ and $B_i$.
		\item In $H$, we have $k^\prime$ $\langle + \rangle$-edge-connected subgraphs $G_1,\cdots,G_t$ with $V(G_i) \supseteq B_i$ for each $i\in[k^\prime]$ and a subgraph $G_{t+1}$ containing only good vertices.
		We use the algorithm \texttt{bad\_cluster}$(G_i,B_i)$ (see Section~3.2) to obtain a clustering $\mathcal{C}_i$ of $G_i$ for each $i\in [k^\prime]$, and use Cohen{-}Addad, Lee and Newman's algorithm~\cite{9996889} to obtain a clustering $\mathcal{C}_{k^\prime+1}$ of $G_{k^\prime+1}$.
		Then the clustering of this partition is set to merging
        $\mathcal{C}_{1}, \cdots, \mathcal{C}_{k^\prime+1}$ together.
			\end{enumerate}
    \item Return the one with the minimum number of mistakes among the clusterings for all partitions of bad vertices in $V^b$. 
    %\item \dots
\end{enumerate}

\subsection{Deal with the Bad Vertices in the Same Cluster}
 
As indicated by Step 2(d) of Algorithm \texttt{CC}, we now need to deal with the following situation: given a graph $\mathbb{G}$ with edges labeled with $\langle + \rangle$ or $\langle - \rangle$ and a set $B$ of bad vertices, the goal is to find a partition of the vertices (clustering) having one subset containing all vertices in $B$ and minimizing disagreements.

To simplify the analysis and handling of bad vertices, we modify the edges incident to bad vertices in $\mathbb{G}$ to construct a new graph $\mathbb{H}$.
For each vertex $v$, we use $l_1,l_2$ to denote the number of $\langle + \rangle,\langle - \rangle$ edges between $v$ and all bad vertices, and  delete $\min\{l_1,l_2\}$ $\langle + \rangle$ and $\langle - \rangle$ edges between $v$ and all bad vertices. 
For more details, refer to Step~1 of Algorithm \texttt{bad\_cluster}.
Then, we partition good vertices into three subsets $N^+(B),N^-(B),N^0(B)$, where
$N^+(B)$ contains the good vertices which have $\langle + \rangle$ edges but no $\langle - \rangle$ edges between them and bad vertices, $N^-(B)$ contains the good vertices which have $\langle - \rangle$ edges but no $\langle + \rangle$ edges between them and bad vertices, and $N^0(B)$ contains the good vertices which are not adjacent to bad vertices.  
The basic idea is that any clustering of $\mathbb{G}$ has to make either $l_1$ or $l_2$ mistakes concerning edges between $v$ and bad vertices.
Thus, it suffices to find a clustering of $\mathbb{H}$, which satisfies the requirement on the mistake number.

Hereby, the primary task is to find a cluster in $\mathbb{H}$ containing all bad vertices, satisfying that the positive and negative mistakes caused by this cluster can be bounded by $O(1) \cdot m_{\mathbb{H}}({\text{OPT}(\mathbb{H})})$, where $\text{OPT}(\mathbb{H})$ is an optimal clustering of $\mathbb{H}$ which partitions bad vertices in the same cluster.
After we have computed this cluster, we can use the polynomial-time $(1.994+ \varepsilon )$-approximation algorithm~\cite{9996889} to find a clustering of the remaining complete graph.
To solve the primary task, we extend the concept of ``$\delta $-clean'' clusters proposed by Bansal et al.~\cite{DBLP:journals/ml/BansalBC04} and distinguish several cases. For some cases, we combine brute-force search and the 3-approximation algorithm by Zuylen and Williamson~\cite{DBLP:journals/mor/ZuylenW09}.

Now, we give a revised definition of the concepts ``$\delta $-good'' and ``$\delta $-clean''.
For~$\mathbb{H}=(V_{\mathbb{H}},E_{\mathbb{H}})$, a good vertex $v$ is called \emph{$\delta$-good} w.r.t. $C\subseteq V_{\mathbb{H}}$ for a positive real-number $\delta$, if  $\vert  N^-(v) \cap C \vert \leq \delta \vert C \vert $ and $\vert  N^+(v) \cap \overline{C} \vert \leq \delta \vert C \vert $.
Since bad vertices are required to be partitioned into one cluster, for a set~$C$ of vertices with $C\supseteq B$, where $B$ is the set of all bad vertices, we say $B$ is \emph{$\delta$-good} w.r.t. $C$, if $\vert  E^-(B, C\setminus B) \vert \leq \delta \vert B \vert \vert C \vert $ and $\vert  E^+(B, \overline{C} \setminus B)\vert \leq \delta \vert B \vert \vert C \vert $. 
Here, $E^-(B, C\setminus B)$ denotes the set of $\langle- \rangle $ edges with one end-vertex in $B$ and the other in $C\setminus B$ and $E^+(B, \overline{C} \setminus B)$ denotes the set of $\langle+ \rangle $ edges with one end-vertex in $B$ and the other in $\overline{C} \setminus B$.
If a good vertex is not $\delta$-good w.r.t $C$, then it is called \emph{$\delta$-bad} w.r.t. $C$.
For a set $C$ of vertices with $C\supseteq B$, the set $B$ is called \emph{$\delta$-bad} w.r.t. $C$ if $B$ is not $\delta$-good w.r.t. $C$.
Finally, for a set $C$ of vertices with $C\cap B=\emptyset$, $C$ is \emph{$\delta$-clean} if all $v\in C$ are $\delta$-good w.r.t. $C$; for a set $C$ of vertices with $C\supseteq B$, $C$ is \emph{$\delta$-clean} if all $v\in C\setminus B$ are $\delta$-good w.r.t $C$ and $B$ is $\delta$-good w.r.t. $C$.

We now give two lemmas, which can be proven in a similar way as Lemmas~4 and 6 in~\cite{DBLP:journals/ml/BansalBC04}.

\begin{lemma}\label{lem:3-mistake}
	Let $C$ be a $\delta$-clean cluster with $C\supseteq B$ and $\delta \leq 1/5$.
    If $\vert C \vert > \frac{1}{\delta}\vert B \vert$, then the number of $\langle - \rangle$ edges whose end-vertices are two good vertices in $C\setminus B$ can be bounded by $2\cdot m_{\mathbb{H}}(\rm{OPT}(\mathbb{H}))$. In addition, if $\vert C \vert > \frac{1}{\delta}\vert B \vert$ and $\vert E^+(B, C\setminus B) \vert \geq \vert E^-(B, C\setminus B) \vert$, then the number of $\langle - \rangle$ edges in $C$ can be bounded by $3\cdot m_{\mathbb{H}}(\rm{OPT}(\mathbb{H}))$.
\end{lemma}

\begin{lemma}\label{lem:optprime}
There exists a clustering $\rm{OPT}^\prime(\mathbb{H})$ of $\mathbb{H}$, in which each non-singleton cluster containing at least one good vertex is $\delta$-clean. In addition, we have $m_{\mathbb{H}}(\rm{OPT}^\prime(\mathbb{H}))\leq$$ (\frac{9}{\delta^2}+1)\cdot m_{\mathbb{H}}(\rm{OPT}(\mathbb{H}))$.
\end{lemma}

From now on, we aim to construct a clustering of $\mathbb{H}$, which, in comparison with the solution $\rm{OPT}^\prime(\mathbb{H})$ of $\mathbb{H}$ given in Lemma~\ref{lem:optprime}, does not make much more mistakes. 
To ease our analysis, we use $C_B^\prime$ to denote the cluster in $\rm{OPT}^\prime(\mathbb{H})$ of, which contains all bad vertices, $C_1^\prime, \cdots, C_t^\prime$ to denote the non-singleton clusters containing only good vertices and $S$ to denote the set of singleton clusters.
We observe that the positive neighbors of bad vertices play an important role in~$C_B^\prime$.
In $C_B^\prime$, these positive neighbors connect bad vertices and the good vertices from $N^-(B)$ and $N^0(B)$.
Thus, $C_B^\prime$ contains some good vertices from $N^-(B)$ and $N^0(B)$, since there are many $\langle + \rangle$ edges between them and the positive neighbors of bad vertices.
In fact, by the definition of $\delta$-good, for $u\in N^+(B)\cap C_B^\prime$, we have that
\begin{align*}
    \vert (N^+[u] \cup B) \cap C_B^\prime \vert &= \vert ( N^+[u] \setminus B) \cap C_B^\prime\vert + \vert B\vert\\ &\geq \vert C_B^\prime \setminus B \vert - \vert N^-(u)\cap C_B^\prime \vert + \vert B \vert \geq (1-\delta) \vert C_B^\prime \vert,\\
    \vert (N^+[u] \cup B) \cap \overline{C_B^\prime} \vert &= \vert N^+(u) \cap \overline{C_B^\prime} \vert \leq \delta \vert C_B^\prime \vert.
\end{align*}
We conclude that for $u\in N^+(B)\cap C_B^\prime$, the set $N^+[u] \cup B$ contains almost all vertices in $C_B^\prime$. 

Next, we give a detailed analysis of $C^\prime_B$, which forms the basis for the case distinction concerning the construction of the cluster containing all bad vertices. If $C'_B$ is large enough, that is, $C'_B$ contains far more good vertices than bad vertices, then the effect of bad vertices is not obvious.
In this case, $C_B^\prime$ behaves like a cluster which consists solely of good vertices and we can apply an algorithm following a similar idea as the one for complete graphs by Bansal et al.~\cite{DBLP:journals/ml/BansalBC04}. 
If the number of vertices in $C_B^\prime$ are bounded, then bad vertices become a significant obstruction.
For this case, we enumerate exhaustively the good vertices that can be added to $C'_B$ and prove that the number of these good vertices is bounded by a function of $|B|$.

In the following, we present Algorithm \texttt{bad\_cluster} which computes, for a given graph $\mathbb{G}$ and a set of bad vertices, a clustering where all bad vertices are put into one cluster and summarizes all above observations. Throughout the rest of this paper, we set $\delta= 1/65$.
\newline

\noindent\textbf{Algorithm \texttt{bad\_cluster}}: Dealing with bad vertices in the same cluster

Input: a graph $\mathbb{G}$ with edge labeling and a set $B$ of bad vertices.

Output: a clustering of $\mathbb{G}$ with a subset containing all bad vertices in $B$.
\begin{enumerate}
    \item Construct a new graph $\mathbb{H}$ from $\mathbb{G}$ as follows.
    Set $e(u,v)=\langle + \rangle$ for all $ u,v \in B ,u\neq v$.
    For all $v \in V(\mathbb{G}) \setminus B$, do the  following: let $l_1,l_2$ denote the number of $+,-$ edges between $v$ and $B$;
    if $l_1 \geq l_2$, then keep arbitrary $l_1-l_2$ $\langle + \rangle$ edges and delete other edges between $v$ and $B$;
    otherwise, keep arbitrary $l_2-l_1$ $\langle - \rangle$ edges and delete other edges between $v$ and $B$.
    \item Use Cohen{-}Addad, Lee and Newman's algorithm~\cite{9996889}  on $\mathbb{H}- B$ to obtain a clustering $\mathcal{C}_1^\prime$.
    Let $\mathcal{C}_2^\prime$ be the clustering by extending $\mathcal{C}_1^\prime$ by one more cluster~$B$. 
    \item In $\mathbb{H}$, we use $N^+(B)$ to denote the positive neighbors of $B$.
    \begin{enumerate}
    \item If $\vert N^+(B) \vert > 2\vert B\vert^2 +\frac{2}{\delta}\vert B\vert $, then for each $v\in N^+(B)$, apply the following algorithm \texttt{clean\_cluster}($\mathbb{H},B, v$) to obtain two clusters $C'(v),C(v)$ containing $v$.\\
    Pick $y\in N^+(B)$ satisfying that $C^\prime(y)\supseteq (B \cup \{y\})$, $\vert C^\prime(y)\vert > \frac{1}{\delta}\vert B\vert$, $C(y)$ is $13\delta$-clean and $\vert E^+(B, \overline{C(y)})\vert \leq \vert E^+(B,\overline{C(v)})\vert$ for all $v \in N^+(B)$.
    \begin{enumerate}
    \item If $\vert E^-(B, C(y)\setminus B)\vert \leq \vert E^+(B, C(y)\setminus B)\vert$, use Cohen{-}Addad, Lee and Newman's algorithm~\cite{9996889}  on $\mathbb{H}- C(y)$ to obtain a clustering $\mathcal{C}^\prime$.
    Merge $C(y)$ and $\mathcal{C}^\prime$ to one clustering and 
    return the one of this clustering and $\mathcal{C}_2^\prime$ with the minimum number of mistakes as output. 
    \item  Else, apply \texttt{more\_negative\_edges}$(\mathbb{H}, B, C(y))$.
    \end{enumerate}
    \item Else, apply \texttt{bounded\_positive\_neighbors}$(\mathbb{H}, B)$. 
    \end{enumerate}

    %\item \dots
\end{enumerate}

In the third step of Algorithm \texttt{bad\_cluster}, we distinguish two cases, the first one corresponding to the case that $C'_B$ is large. Hereby, we need the following algorithm to computer two clusters.
If there is a good vertex $u\in N^+(B)\cap C_B^\prime$, we iteratively ``clean'' $N^+[u] \cup B$ as follows.
\newline

\noindent\textbf{Algorithm \texttt{clean\_cluster}}: producing two clusters.

Input: a graph $\mathbb{H}$, a set $B$ of bad vertices and a good vertex $u$.

Output: two clusters $C^\prime$ and $C$ containing $u$. 
\begin{enumerate}
    \item Let $A(u)=N^+[u]\cup B$ with $N^+[u]=N^+(u)\cup \{u\}$.
    \item (Vertex Removal Step):  While $\exists v \in A(u)\setminus\{u\}$ such that $v$ is $3\delta$-bad w.r.t.$A(u)$, then set $A(u)=A(u)\setminus \{v\}$.\\
    If $B$ is $3\delta$-good w.r.t. $A(u)$ and $u$ is $3\delta$-good w.r.t. $A(u)$, then set $C^\prime=A(u)$;
    otherwise, set $C'=\{u\},C=\{u\}$ and return $C',C$.
    \item (Vertex Addition Step): Let $V^\prime=\{v\in V(\mathbb{H}): \vert N^+(v) \cap A(u)\vert \geq (1-9 \delta)\vert A(u)\vert, \vert N^+(v)\cap \overline{A(u)}\vert \leq 9 \delta \vert A(u)\vert  \}$. Set $C=A(u)\cup V^\prime$.
    \item Return $C^\prime,C$.
    %\item \dots
\end{enumerate}

The next two lemmas connect $\vert N^+(B) \vert$ with $\vert C'_B \vert$. 
We first observe that since $C_B^\prime$ is $\delta$-clean and $C_B^\prime \supset B$, if there are many positive neighbors of bad vertices, then $C_B^\prime$ contains a large portion of positive neighbors of bad vertices, and thus, $C_B^\prime$ is very large.
The following lemma provides a formal description of this observation.

\begin{lemma}\label{lem:largecase}
	If $\vert N^+(B) \vert > 2\vert B\vert^2 +\frac{2}{\delta}\vert B\vert $, then $\vert C_B^\prime \vert > \frac{2}{\delta}\vert B\vert$.
\end{lemma}

The next lemma establishes the connection between $|C'_B|$ and the existence of $\delta$-good vertices.
Recall that  we use $C_1^\prime, \cdots, C_t^\prime$ to denote the non-singleton clusters containing only good vertices and $S$ to denote the set of singleton clusters in $\rm{OPT}^\prime(\mathbb{H})$ given in Lemma~\ref{lem:optprime}.

\begin{lemma}\label{lem:existx}
	If $\vert C_B^\prime \vert > \frac{2}{\delta}\vert B\vert$, then for every $x\in N^+(B)\cap C_B^\prime$, we have that $C^\prime(x) \supseteq (B\cup \{x\}), \vert C^\prime(x) \vert > \frac{1}{\delta}\vert B\vert$, $C(x)$ is $13\delta$-clean, $C(x)\supseteq C_B^\prime$ and $C(x)\cap C_i^\prime =\emptyset$ for each $i\in[t]$, where $C^\prime(x)$ and $C(x)$ are the output of Algorithm \rm{\texttt{clean\_cluster}}$(\mathbb{H},$ $B,x)$.
\end{lemma}
 
Lemma~\ref{lem:existx} guarantees that if $N^+(B)$ is large enough, then Step 3(a) of Algorithm \texttt{bad\_cluster} can find a vertex $y\in N^+(B)$, such that $C^\prime(y)$ and $C(y)$ have the required properties.
And these properties lead to only three cases of $C(y)$ we need to consider, as shown by the following lemma. 
Thus, we can bound the number of $\langle + \rangle$ edges between $C(y)$ and other clusters by Lemma~\ref{lem:optprime} and the property $\vert E^+(B,\overline{C(y)})\vert \leq \vert E^+(B, \overline{C(v)})\vert$ for all $v \in N^+(B)$.

\begin{lemma}\label{lem:3case}
	The cluster $C(y)$ computed at Step 3(a) of {\rm{\texttt{bad\_cluster}}}  satisfies one of the following three conditions: \\
    1. $(C(y) \setminus B)
    \subseteq S $;\\
    2. $C(y) \supseteq C_B^\prime$ and $C(y)\cap C_i^\prime= \emptyset$ for any $i\in[t]$;\\
    3. $C(y) \supseteq C_j^\prime, C(y) \cap C_B^\prime= B$, and $C(y)\cap C_i^\prime= \emptyset$, for any $i\neq j, i\in[t]$.
\end{lemma}

If $C(y)$ satisfies Condition 1 or 3, we output $B$ as a cluster, corresponding to the clustering $\mathcal{C}_2^\prime$ in \texttt{bad\_cluster}.
If $C(y)$ satisfies Condition 2, we find a cluster $C(y)$ satisfying $C(y) \supseteq C_B^\prime$, $C(y)$ being $13\delta$-clean and $\vert C(y)\vert > \frac{1}{\delta}\vert B\vert$.
If $\vert E^-(B, C(y)\setminus B)\vert \leq \vert E^+(B, C(y)\setminus B)\vert$, we output $C(y)$ as a cluster according to Lemmas~\ref{lem:3-mistake} and \ref{lem:optprime}; 
otherwise, we apply \texttt{more\_negative\_edges} shown in the following. 

The case $\vert E^-(B, C(y)\setminus B)\vert > \vert E^+(B, C(y)\setminus B)\vert$ occurs, when there are too many negative neighbors of bad vertices in $C(y)$.
If the number of these negative neighbors can be bounded by $3\vert B \vert$, then we can guess the negative neighbors which are not in  $C_B^\prime$ and remove them, in Step 2 of \texttt{more\_negative\_edges};
otherwise, we pick arbitrary $3\vert B \vert$ many such negative neighbors and consider two cases, one being at least $|B|$ of them in $C'_B$ and the other at least $2|B|$ such negative neighbors not in $C'_B$.
The former case, we output $B$ as a cluster, while the latter is coped with \texttt{good\_replace\_bad} shown later.
\newline

\noindent\textbf{Algorithm \texttt{more\_negative\_edges}}: dealing with the cluster $C$ with $\vert E^-(B, C\setminus B ) \vert > \vert E^+(B, C\setminus B)\vert$

Input: a graph $\mathbb{H}$, a set $B$ of bad vertices, a cluster $C$ containing $B$.

Output: a clustering of $\mathbb{H}$ with a subset containing all bad vertices in $B$.
\begin{enumerate}
    \item Use Cohen{-}Addad, Lee and Newman's algorithm~\cite{9996889}  on $\mathbb{H}- B$ to obtain a clustering $\mathcal{C}_1^\prime$, and then let $\mathcal{C}_2^\prime$ be the clustering resulted by merging $\mathcal{C}_1^\prime$ and~$B$.
    \item If $\vert N^-(B)\cap C\vert < 3\vert B \vert$, 
    \begin{enumerate}
    \item For each subset $R$ of $N^-(B)\cap C$, use Cohen{-}Addad, Lee and Newman's algorithm~\cite{9996889}  on $\mathbb{H}- (C\setminus R)$ to obtain a clustering $\mathcal{C}_3^\prime$. The clustering of this subset is set to merging $\mathcal{C}_3^\prime$ and $C\setminus R$.
    \item Return the one clustering for all subsets and $\mathcal{C}_2^\prime$, which has the minimum number of mistakes. 
    \end{enumerate}

    \item Else, pick arbitrarily $3 \vert B \vert $ vertices in $N^-(B)\cap C$, put them into the set $N$.
    \begin{enumerate}
    \item For each subset $B^-$ of $N$, if $\vert B^-\vert \geq \vert B \vert$, 
    the clustering of this subset is set to the output of \texttt{good\_replace\_bad}($\mathbb{H},B,B^+$).
    \item Return the one clustering for all subsets of $N$ and $\mathcal{C}_2^\prime$, which has the minimum number of mistakes. 
    \end{enumerate}

    %\item \dots
\end{enumerate}

In Algorithm \texttt{good\_replace\_bad}, we are given at least $\vert B \vert$ good vertices that are required to be  partitioned into the same cluster as bad vertices in $C_B^\prime$; in other words, these good vertices are assumed to be partitioned into the same cluster as bad vertices in an optimal clustering OPT($\mathbb{H}$).
We use $B^+$ to denote these good vertices.
We replace the bad vertices by these good vertices in another auxiliary graph $\hat{\mathbb{H}} $.
The construction of $\hat{\mathbb{H}} $ is given in \texttt{good\_replace\_bad}.
The basic idea is that the number of $\langle + \rangle$ edges in  $\hat{\mathbb{H}} $ between a good vertex and $B^+$ is at least half the number of $\langle + \rangle$ edges in $\mathbb{H}$ between the good vertex and $B^+ \cup B$, and the same holds for $\langle - \rangle$ edges.
Thus, the mistakes in $\hat{\mathbb{H}} $ of any clustering is at least half the mistakes in $\mathbb{H}$ of the same clustering.
Finally, we apply Zuylen and Williamson's algorithm~\cite{DBLP:journals/mor/ZuylenW09} to the complete graph $\hat{\mathbb{H}} $ to produce a clustering satisfying that the good vertices in $B^+$ are in the same cluster.
\newline

\noindent\textbf{Algorithm \texttt{good\_replace\_bad}}: dealing with at least $\vert B \vert$ good vertices which need to be in the same cluster as bad vertices 

Input: a graph $\mathbb{H}$, a set $B$ of bad vertices, a set $B^+$ of good vertices with $\vert B^+ \vert \geq \vert B \vert$.

Output: a clustering of $\mathbb{H}$ with a subset containing all vertices in $B \cup B^+$.

\begin{enumerate}
    \item Construct a new graph $\hat{\mathbb{H}} $ from $\mathbb{H}$ as follows.
    \begin{enumerate}
    \item For all $v \in V(\mathbb{H}) \setminus (B\cup B^+)$, let $l_3$ and $l_4$ denote the numbers of $\langle + \rangle$ and $\langle - \rangle$ edges between $v$ and $B$ in $\mathbb{H}$, respectively, 
    let $l_5$ and $l_6$ denote the numbers of $\langle + \rangle$ and $\langle - \rangle$ edges  between $v$ and $B^+$ in $\mathbb{H}$, respectively. 
    \begin{enumerate}
        \item If $l_3 > l_5$, then change arbitrary $\frac{l_6}{2}$  $\langle - \rangle$ edges between $v$ and $B^+$ to $\langle + \rangle$ edges.
    \item Else, if $l_4 >l_6$, then change arbitrary $\frac{l_5}{2}$ $\langle + \rangle$ edges  between $v$ and $B^+$ to $\langle - \rangle$ edges.
    \end{enumerate}
    \item Afterwards, remove all the bad vertices and the edges incident to them to obtain  $\hat{\mathbb{H}} $.
    \end{enumerate}
    \item Apply Zuylen and Williamson's algorithm~\cite{DBLP:journals/mor/ZuylenW09} to $\hat{\mathbb{H}} $ with $P^+=\{(u,v)\vert u,v\in B^+, u\neq v\}$ and $P^-= \emptyset $ to get a clustering $\mathcal{C}_3^\prime $ satisfying that there exists a cluster $C^\prime_3 \supseteq B^+$ in  $\mathcal{C}_3^\prime$.
    \item Return the clustering resulted by extending the cluster $C^\prime_3$ in $\mathcal{C}_3^\prime$ by adding $B$ to it.
    %\item \dots
\end{enumerate}

As the last part of \texttt{bad\_cluster}, \texttt{bounded\_positive\_neighbors} deals with bad vertices satisfying $\vert N^+(B) \vert \leq 2\vert B\vert^2 +\frac{2}{\delta}\vert B\vert $.
We use $B^+$ to denote the good vertices in $C_B^\prime$ which are connected to bad vertices by positive edges or paths consisting  only of positive edges.
In Step~1 of \texttt{bounded\_positive\_neighbors}, we guess the positive neighbors of bad vertices in $C_B^\prime$ and apply Procedure \texttt{find\_neighbors} to deal with distinct cases of these neighbors.
\newline
%If there are no positive neighbors of bad vertices in $C_B^\prime$, then we output $B$ as a cluster.
%Otherwise, we use \texttt{good\_replace\_bad} to deal with the case that there are at least $\vert B \vert $ good vertices in $B^+_1$.
%For $1 \leq \vert B^+_1 \vert < \vert B \vert$, we iteratively guess the positive neighbors of $B^+_j$ in $C_B^\prime$ and add them to $B^+_{j+1}$.
%Note that in each For-loop of \texttt{bounded\_positive\_neighbors}, either the enumeration case is over or at least one new vertex is found. 

\noindent\textbf{Algorithm \texttt{bounded\_positive\_neighbors}}: dealing with bad vertices satisfying  $\vert N^+(B) \vert \leq 2\vert B\vert^2 +\frac{2}{\delta}\vert B\vert $. 

Input: a graph $\mathbb{H}$ with edge labeling, a set $B$ of bad vertices.

Output: a clustering of $\mathbb{H}$ with a subset containing all vertices in $B$.

\begin{enumerate}
    \item For each subset $N$ of $N^+(B)$, do as follows.
    \begin{enumerate}
    \item If $N \neq \emptyset$, pick an arbitrary vertex $u\in N$, run \texttt{clean\_cluster}$(\mathbb{H}, B, u)$ to obtain two clusters $C^\prime(u), C(u)$ containing $u$.
    \begin{enumerate}
    \item If $\vert E^-(B, C(u)\setminus B)\vert \leq \vert E^+(B, C(u)\setminus B)\vert$, use Cohen{-}Addad, Lee and Newman's algorithm~\cite{9996889} on $\mathbb{H}- (C(u)\cup B)$ to obtain a clustering $\mathcal{C}^\prime$, and set $\mathcal{C}^{\prime \prime}$ be the clustering by merging $C(u)\cup B$ and $\mathcal{C}^\prime$.
    \item Else, set $\mathcal{C}^{\prime \prime}=$  \texttt{more\_negative\_edges}$(\mathbb{H},B, C(u)\cup B)$.
    \end{enumerate}
        \item Apply Procedure \texttt{find\_neighbors}($\mathbb{H}$,$B$,$\emptyset$,$N$,$\mathcal{C}^{\prime \prime}$).

    \end{enumerate}
    \item Return the one clustering for all subsets which has the minimum number of mistakes. 
\end{enumerate}
In Procedure \texttt{find\_neighbors}, if Cases 1-3 occur, then Steps 2-4 compute the clustering for the corresponding subset and terminate; otherwise, we have $|B^+| < |B|$. Then Step~5 iteratively applies \texttt{find\_neighbors} to
guess the positive neighbors of $B^+$ in $C_B^\prime$, which need to be added to $B^+$.
In Step 5, there are at most $2^{O(|B|^2)}$ subsets of $N^+(B^+)$.
For each subset, \texttt{find\_neighbors} can be iteratively applied at most $|B|$ times, since each time when \texttt{find\_neighbors} is applied, either the enumeration case is over (Steps~2-4) or at least one new vertex is found (Step~5). 
Therefore, \texttt{find\_neighbors} iterates at most $2^{O(|B|^3)}\cdot {\rm{poly}}(n)$ times. 
\newline

\noindent\textbf{Procedure \texttt{find\_neighbors}}

Input: a graph $\mathbb{H}$ with edge labeling, a set $B$ of bad vertices,  a set $B^+$ of good vertices, a set $N^\prime$ of good vertices, a clustering $\mathcal{C}^{\prime \prime}$ of $\mathbb{H}$.

\begin{enumerate}
    \item Set $B^+=B^+\cup N^\prime$ and use $N^+(B^+)$ to denote the positive neighbors of $B^+$ in the graph $\mathbb{H}- (B^+ \cup B)$.
    \item Case 1: if $N^\prime=\emptyset$, then use Cohen{-}Addad, Lee and Newman's algorithm~\cite{9996889}  on $\mathbb{H}- (B\cup B^+)$ to obtain a clustering $\mathcal{C}_1^\prime$. The clustering for this subset $N^\prime$ is set to merging $B^+\cup B$ and $ \mathcal{C}_1^\prime$. 
    \item Case 2: if $\vert B^+ \vert \geq \vert B \vert$, then the clustering for this subset $N^\prime$ is set to be the result of \texttt{good\_replace\_bad}$(\mathbb{H}, B, B^+)$.
    \item Case 3: if $\vert N^+(B^+)\vert > (2+\frac{2}{\delta})\vert B \vert ^2$, 
    then the clustering for this subset $N^\prime$ is set to be $\mathcal{C}^{\prime\prime}$.
    \item Case 4: if $N^\prime\neq \emptyset$, $\vert B^+ \vert < \vert B \vert$ and $\vert N^+(B^+)\vert \leq (2+\frac{2}{\delta})\vert B \vert ^2  $, then
    for each subset $N^{\prime\prime}$ of $N^+(B^+)$, apply \texttt{find\_neighbors}($\mathbb{H}$,$B$,$B^+$,$N^{\prime\prime}$,$\mathcal{C}^{\prime \prime}$).

\end{enumerate}

We arrive at our main result.
 
\begin{theorem}
    {\rm{Algorithm \texttt{CC}}} returns in $2^{O(k^3)} \cdot {\rm{poly}}(n)$ time a $(\frac{18}{\delta^2}+7.3)$-factor approximation for Correlation Cluster on general graphs, where $\delta =\frac{1}{65}$ and $k$ is the minimum number of vertices whose removal results in a complete graph.
\end{theorem}

\section{Conclusion}
Based on the concept of ``distance from approximability'', we present the first FPT constant-factor approximation algorithm for Correlation Clustering on general graphs, parameterized by the minimum number of vertices whose removal results in a complete graph.
In comparison with the $(1.994+ \varepsilon )$-approximation~\cite{9996889} for complete graphs, it is crucial to improve the constant approximation factor, which is currently 76057.3.
A combination with randomized algorithms or LP relaxations similarly as~\cite{10.1145/1411509.1411513,CHARIKAR2005360,10.1145/2746539.2746604,9996889,DBLP:journals/mor/ZuylenW09} seems a promising approach.
Moreover, it is interesting to consider other parameters, for instance, vertex-deletion distance to a cluster graph (a graph in which each connected component is a clique).  
Finally, based on the success on TSP~\cite{zhou_et_al:LIPIcs.ISAAC.2022.50} and Correlation Clustering, ``distance from approximability'' might be applicable to other optimization problems.

\section{Acknowledgements}
The authors are grateful to the reviewers of COCOON 2024 for their valuable comments and constructive suggestions.

%
% ---- Bibliography ----
%
% BibTeX users should specify bibliography style 'splncs04'.
% References will then be sorted and formatted in the correct style.
%
\bibliographystyle{splncs04}
% \bibliography{mybibliography}
\bibliography{references}

\end{document}